\documentclass[aps,pra,superscriptaddress, twocolumn,longbibliography]{revtex4-2}

\usepackage[english]{babel}
\usepackage{times}
\usepackage{graphicx}
\usepackage{graphics}
\usepackage{amsmath}
\usepackage{xcolor}
\usepackage{mathtools}
\usepackage{amsfonts}
\usepackage{amssymb}
\usepackage{dsfont}
\usepackage{epstopdf}
\usepackage{makeidx}
\usepackage{subfigure}
\usepackage{color}
\usepackage{pgf}
\usepackage{bm}
\usepackage{ulem}
\usepackage{hyperref}
\usepackage{orcidlink}

\makeindex

\begin{document}

\title{Quantum Impurities as Probes of Finite-Temperature Fluctuations in  Two-Dimensional Bose Gases}

\author{Victor Velasco}
\email{vvelasco@sissa.it}
\affiliation{International School for Advanced Studies (SISSA), Via Bonomea 265, I-34136 Trieste, Italy}

\author{Gabriele Spada}

\affiliation{CINECA Consorzio Interuniversitario, Via Magnanelli 6/3, 40033 Casalecchio di Reno, Italy}

\affiliation{School of Science and Technology, Physics Division, Università di Camerino, 62032 Camerino, Italy}

\author{Giovanni Midei}

\affiliation{School of Science and Technology, Physics Division, Università di Camerino, 62032 Camerino, Italy}

\affiliation{Pitaevskii BEC Center, CNR-INO, Trento, Italy}

\affiliation{INFN, Sezione di Perugia, I-06123 Perugia, Italy}

\author{Andrea Perali} 

\affiliation{School of Pharmacy, Physics Unit, University of Camerino, Via Madonna delle Carceri 9, 62032 Camerino, Italy}

\author{Luis A. Peña Ardila}
\email{luis.ardila@units.it \\ luis.ardila@ts.infn.it}
\affiliation{Dipartimento di Fisica, Università di Trieste, Strada Costiera 11, I-34151 Trieste, Italy}

\affiliation{
Istituto Nazionale di Fisica Nucleare (INFN), 
Trieste Section, Via Valerio 2, I-34127 Trieste, Italy
}

\begin{abstract}
Two-dimensional quantum gases provide a distinctive setting in which enhanced thermal fluctuations, finite-size effects, and two-body bound-state formation are intrinsically intertwined. In this work, we study a single attractive impurity immersed in a finite, weakly interacting two-dimensional Bose gas, where finite size stabilizes a nonzero condensate fraction by introducing an infrared momentum scale, thereby enabling a Bogoliubov description of the bath. Using a hybrid approach that combines finite-temperature many-body scattering theory with input from path-integral Monte Carlo, we analyze the impurity quasiparticle energy across the condensate and normal regimes. The infrared scale generates a phonon-activation temperature below which the impurity energy remains nearly temperature independent. Once the resolved phonon modes become thermally populated, their contribution competes with condensate depletion, producing a nonmonotonic temperature dependence of the polaron energy. These results suggest that attractive Bose polarons may serve as sensitive probes of finite-size thermal fluctuations, phonon dressing, and bound-state physics in low-dimensional Bose gases.
\end{abstract}

\maketitle

\section{Introduction}

The interaction of a mobile particle with a quantum many-body system provides a general framework for investigating quasiparticle formation, medium-induced dressing, and emergent correlations. The paradigmatic example is the polaron, originally introduced by Landau to describe an electron dressed by the lattice polarization that it induces in an ionic crystal~\cite{Landau1933}, and later developed through Feynman’s path-integral treatment of the Fr\"ohlich model~\cite{Feynman1955}. More broadly, problems of a mobile impurity embedded in an interacting medium provide a versatile framework for understanding how the properties of a single particle are renormalized through its coupling to the low-energy excitations of the medium. Ultracold atomic gases provide a highly controllable platform for realizing such impurity problems, since dimensionality, density, temperature, and interaction strengths can be tuned experimentally \cite{Chin2010}. This has enabled systematic studies of impurities immersed in both fermionic~\cite{Schirotzek2009,Nascimbene2009,Marco2012,Kohstall2012,Ong2015,Yi2015,Cetina2015,Cetina2016,Scazza2017} and bosonic reservoirs~\cite{PenaArdila2015Impurity,PenaArdila2019_2,Jorgensen2016,Hu2016,Grusdt2017,Yoshida2018,StrongArdilla,Yan2019,Pastukhov2018,Pastukhov2020,Skou2021,Parish2024,Grusdt2025Impurities}, giving rise to Fermi and Bose polarons, respectively.\\

In the three-dimensional (3D) Bose polaron case, at low temperature, the impurity is dressed by Bogoliubov excitations of the bosonic condensate, while at finite temperature it also interacts with thermally populated modes. The impurity response therefore contains information about condensate depletion, thermal fluctuations, quasiparticle damping, and the redistribution of spectral weight~\cite{Liu2019}. Previous studies have shown that finite temperature can substantially modify the Bose polaron energy, residue, and linewidth, and that the impurity can act as a probe of the state of the Bose gas near phase transitions ~\cite{Jesper2017,PenaArdila2019_2, Guenther2018, Dzsotjan2020, Field2020, Pascual2021, Brunn2022, Amelio2024, Drescher2024}.

While higher-dimensional systems have been studied extensively, the two-dimensional (2D) case remain comparatively less explored despite exhibiting potentially richer many-body physics. In the thermodynamic limit, true Bose--Einstein condensation is forbidden at any finite temperature by the Mermin--Wagner--Hohenberg theorem~\cite{Mermin1966,Hohenberg1967}. Nevertheless, superfluidity can persist through the Berezinskii--Kosterlitz--Thouless (BKT) mechanism, where bound vortex--antivortex pairs preserve quasi-long-range order below the transition, while above it vortex unbinding destroys phase coherence and leads to the universal jump of the superfluid density~\cite{Berezinskii1971,Berezinskii1972,Kosterlitz1973,Kosterlitz1974,Nelson1977}. Thus, in 2D, finite-temperature impurity physics is inseparably connected to enhanced phase fluctuations, thermal phonons, and vortex physics as well as the physics of finite-size thermal fluctuations.

Experimentally realized two-dimensional Bose gases are necessarily finite. Confining traps, box geometries, and finite observation regions introduce an effective infrared momentum cutoff, $k_{\min}$, thereby suppressing fluctuations with wavelengths comparable to or larger than the characteristic system size. As a consequence, a finite system can support a nonzero condensate fraction at low temperature, even though true long-range order is absent in the thermodynamic limit. The temperature range over which the finite-size condensate is depleted is expected to overlap with, or occur near, the finite-size crossover associated with the BKT transition, in agreement with functional renormalization-group analyses, Monte Carlo simulations, and experimental observations~\cite{Floerchinger2009,Prokofev2001,Prokofev2002,Hadzibabic2006,Clade2009,Hung2011}. This finite-size condensate provides a controlled Bogoliubov background for impurity dressing, while the infrared cutoff introduces an additional physical scale that can strongly affect the low-energy impurity response. 

Despite this relevance, finite-temperature Bose polarons in 2D Bose gases remain less explored than their 3D counterparts. Existing theoretical work has addressed the problem using, for example, the functional renormalization group ~\cite{Felipe2024} and stochastic projected Gross--Pitaevskii simulations~\cite{Amelio2024}, while experimental studies are still scarce~\cite{Schlederer2024}. In particular, recent work has focused on repulsive impurities as spectroscopic probes of vortex proliferation across the BKT transition~\cite{Amelio2024}. In that case, the bosonic density depletion inside vortex cores creates attractive potential wells for the impurity and impurity--vortex binding becomes a central mechanism. On the other hand, the attractive Bose polaron in a finite 2D gas provides a complementary scenario that probes condensate depletion, thermally activated phonon dressing, and the intrinsic 2D scattering bound state scale.\\

In this work, we address this complementary regime by investigating an attractive mobile impurity immersed in a 2D Bose gas, with particular emphasis on the interplay between quantum and thermal fluctuations. The finite size introduces an infrared momentum cutoff that stabilizes a nonzero condensate fraction at low temperature, which we determine using path-integral Monte Carlo (PIMC) simulations. This allows us to formulate a Bogoliubov description of the bath in the condensed phase. Using an extended non-self-consistent \(T\)-matrix (e-NSCT) approach, adapted closely from the finite-temperature 3D formulation~\cite{Guenther2018}, we show that the infrared cutoff generates a finite-size phonon-activation window. In this window, thermally activated phonon dressing competes with condensate depletion, producing a nontrivial temperature dependence of the attractive polaron energy. These finite-size infrared effects establish attractive Bose polarons as spectroscopic probes of finite-temperature fluctuations in 2D Bose gases.\\

The paper is organized as follows. We introduce the finite-temperature 2D Bose gas  and its relevant energy scales in Sec.~\ref{sec:bosegas}.  Section~\ref{sec:phonon_activation_window} identifies  the finite-size phonon-activation window and characterizes  the momentum scales that control the thermal phonon dressing of the impurity.  In Sec.~\ref{sec:polaron} we introduce the impurity  Hamiltonian and develop both the perturbative Fr\"ohlich treatment, valid at weak coupling, and the non-perturbative  e-NSCT scheme for the strong-coupling regime.  Section~\ref{Sec: Temp_effects} presents and discusses  our main results for the polaron energy as a function of temperature and impurity--boson coupling. Experimental signatures and feasibility estimates are  collected in Sec.~\ref{Sec: experimental}. Finally, Sec.~\ref{Sec: conclusions} summarizes our findings and outlines future directions.

\section{Finite temperature 2D Bose gas}
\label{sec:bosegas}
Before addressing the impurity physics, we first introduce the relevant energy scales of the weakly interacting 2D Bose gas. The Hamiltonian describing this system can be written as
\begin{align}
    \hat{H}_B = \sum_{\mathbf{k}}\varepsilon_{\mathbf{k}}\, \hat{a}^{\dagger}_{\mathbf{k}}\hat{a}_{\mathbf{k}} + \frac{g_{BB}}{2\mathcal{A}}\sum_{\mathbf{k},\mathbf{k}',\mathbf{q}}
    \hat{a}^{\dagger}_{\mathbf{k}+\mathbf{q}}\hat{a}^{\dagger}_{\mathbf{k}'-\mathbf{q}}
    \hat{a}_{\mathbf{k}'}\hat{a}_{\mathbf{k}},
\end{align}
where $\hat{a}_\mathbf{k}$ and $\hat{a}^{\dagger}_\mathbf{k}$ are bosonic annihilation and creation operators and $\mathcal{A}=L^2$ is the system area, defined by  $L=\sqrt{N/n}$, where $N$ is the total number of bosons and $n$ the total density. The first term corresponds to the kinetic energy of the bosons, with single-particle dispersion $\varepsilon_{\mathbf{k}} = k^2/2m$, and the second term describes their interaction, modeled by a contact potential with strength $g_{BB}$. We use natural units $k_{\mathrm{B}}=\hbar=1$.

This finite 2D system can stabilize a nonzero condensate fraction at a nonzero temperature. This enables the use of Bogoliubov theory for the bosonic sector. Diagonalizing the interacting Bose gas in terms of Bogoliubov quasiparticles yields $\hat{H}_{\mathrm{Bogo}} = \sum_{\mathbf{k}} E_{\mathbf{k}}\, \hat{b}^{\dagger}_{\mathbf{k}}\hat{b}_{\mathbf{k}},$ where $\hat{b}_{\mathbf{k}}$ and $\hat{b}^\dagger_{\mathbf k}$ are Bogoliubov operators. This Bogoliubov description should be understood as a low-energy theory, and its linear phonon limit applies only at momenta below the healing scale. The excitation spectrum is written as $E_{\mathbf{k}} = \sqrt{\,\varepsilon_{\mathbf{k}}\left[\varepsilon_{\mathbf{k}} + 2 n_0(T) g_{BB}\right]\,},$ which is the dispersion of Bogoliubov excitations, with $n_0(T)$ the temperature-dependent condensate density.

\begin{figure}[!t]
    \centering
    \includegraphics[width = \linewidth]{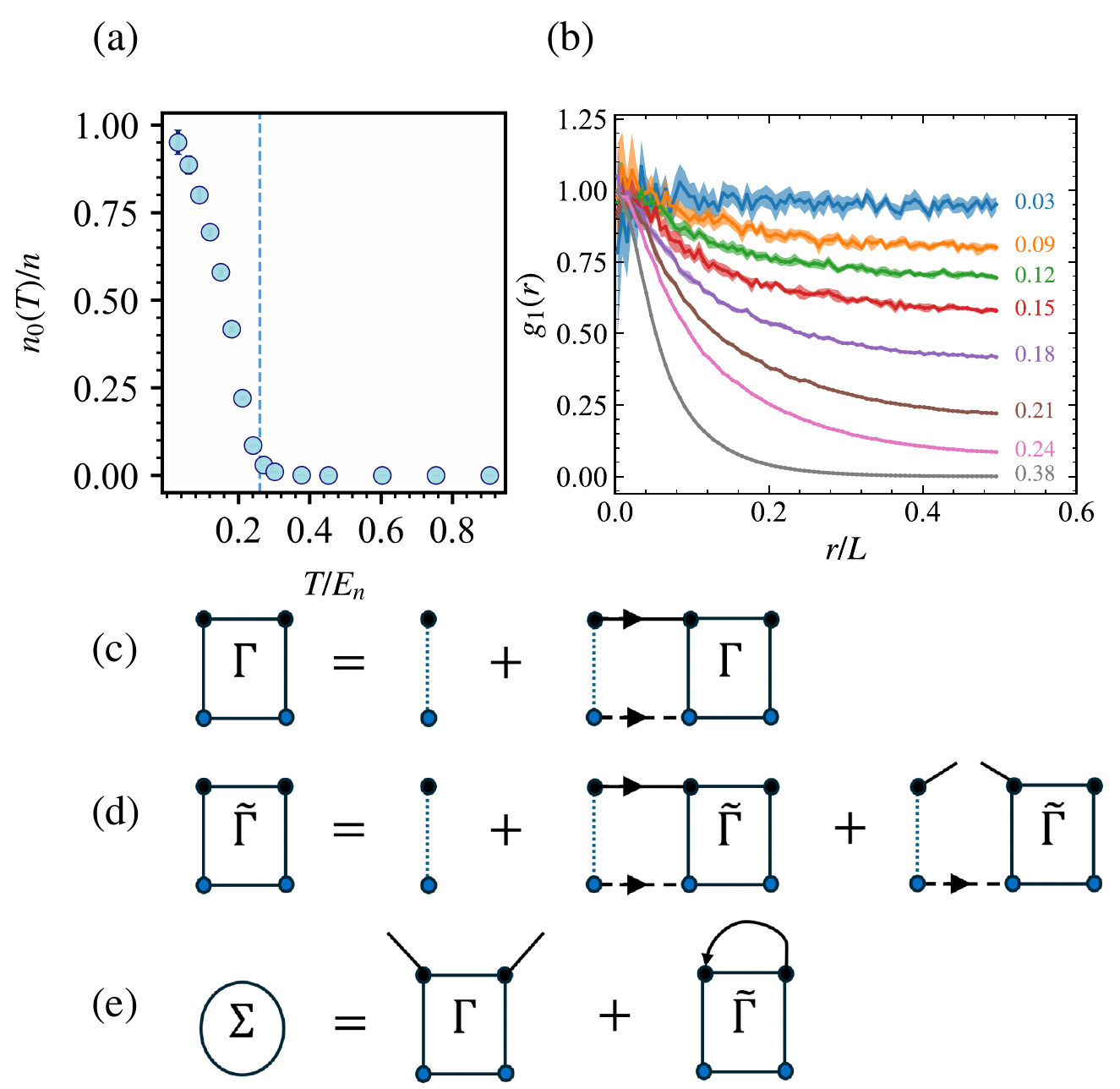}
    \caption{(a) Condensate fraction $n_{0}(T)/n$ as a function of temperature calculated using PIMC for a repulsive boson--boson interaction ($g_{BB}=0.5\hbar^2/m$) and $N = 512$ particles. The vertical dashed line represents $\Theta_c$. (b) One-body correlation functions, $g_1(r)$, obtained from PIMC simulations. The shaded area around the curves represents the statistical error. The temperature values, in units of $E_n$, are indicated by the color scale, and the condensate fraction is extracted from $g_1(L/2)$. 
(c)-(e) Feynman diagrams representing the nonextended and extended schemes for determining the
self-energy \cite{Guenther2018}. Solid and dashed lines denote bosonic and impurity propagators,
respectively, while the dashed vertical line represents the two-body interaction.}
    \label{fig: 1}
\end{figure}

In order to determine $n_{0}(T)$, we employ path-integral Monte Carlo calculations for a weakly interacting Bose gas in a box with $N=512$ particles at fixed total density $n$ and coupling strength $g_{BB} = 0.5\,\hbar^2/m$ (see \hyperref[app:appendix A]{Appendix A} for details). The finite system size introduces a natural IR momentum cutoff $k_{\min}$, which follows directly from the linear system size $L$. In two dimensions, the relation $n = N/L^2$ defines $L$, and therefore the minimal momentum is given by
\begin{equation}
k_{\min} = \frac{2\pi}{L} = 2\pi \sqrt{\frac{n}{N}}.
\end{equation}
For the parameters considered in this work, the finite-size infrared cutoff is
$k_{\min}=7.8\times10^{-2}k_n$, where
$k_n=\sqrt{4\pi n}$ denotes the characteristic momentum scale associated with the total density $n$. We use the corresponding density energy scale $E_n=k_n^2/2m$ to express energies throughout. We estimate the finite-size condensate density $n_0(T)$ from the long-distance value of the one-body correlation function $g_1(r)$, evaluated at the maximum independent separation $r=L/2$. The resulting condensate estimator is shown in Fig.~\ref{fig: 1}(a), while the corresponding correlation functions are shown in Fig.~\ref{fig: 1}(b). As the temperature increases, the finite-size condensate fraction decreases smoothly, reflecting the progressive loss of long-distance coherence in the temperature region where BKT-related fluctuations become important \cite{Floerchinger2009,Prokofev2001,Prokofev2002,Hadzibabic2006,Clade2009,Hung2011}.\\

In terms of the IR cutoff, the Bogoliubov description is controlled only while the condensate interaction scale, \(g_{BB}n_0(T)\), remains larger than the minimum kinetic energy imposed by the infrared cutoff, \(\varepsilon_{\min}=k_{\min}^2/2m\). Physically, this ensures that the lowest accessible modes are still interaction dominated and retain their collective phononic character. We therefore define a size-dependent characteristic temperature, \(\Theta_c\), through the condition $g_{BB}n_0(\Theta_c)=\varepsilon_{\min}$. Using the PIMC data, this condition gives \(\Theta_c/E_n=0.246\). This scale should not be regarded as the thermodynamic transition temperature. Rather, \(\Theta_c\) denotes the finite-size Bogoliubov-validity boundary of the present calculation. Equivalently, this condition corresponds to the healing length becoming comparable to the infrared length \(1/k_{\min}=L/(2\pi)\), which is proportional to interparticle distance, $\ell = n^{-1/2}$. This interpretation is consistent with finite-size treatments in which the infrared flow is stopped at a momentum scale set by the inverse system size, \(L^{-1}\)~\cite{Floerchinger2009}, and in which the condensate density is a finite-size quantity controlled by the IR cutoff~\cite{Holzmann2007}. Therefore, $\Theta_c$ identifies a finite-size  crossover region in which Bogoliubov theory gradually loses its validity, rather than a sharp phase transition. Below this scale, the bath can be treated as a condensate supporting collective phononic excitations within Bogoliubov theory. Above $\Theta_c$, we model the bath by a weakly interacting normal Bose gas at the Hartree mean-field level, where the boson--boson interaction produces a rigid shift $2g_{BB}n$ of the particle-like spectrum \cite{Uchino2022,Hilker2022}. The excitations are therefore described by $E_{\mathbf k}=\varepsilon_{\mathbf k}+2g_{BB}n-\mu(T)
=\varepsilon_{\mathbf k}-\mu_B(T)$, where $\mu_B(T)\equiv\mu(T)-2g_{BB}n$ is the Hartree-shifted effective chemical potential, fixed by the finite-size density equation 
\begin{equation}
n  = \int_{k_{\min}}^\infty \frac{d^2k}{(2\pi)^2}\frac{1}{e^{(\varepsilon_{\mathbf{k}}-\mu_B)/T}-1}.
\label{eq:n-2D-cutoff}
\end{equation}
For the finite-size system considered here, we replace the discrete box sum by a continuum integral with the lower cutoff $k_{\min}=2\pi/L$ (see \hyperref[app:appendix C]{Appendix C} for details), which gives an analytical expression for the effective chemical potential,
\begin{equation}
\mu_B(T)=\varepsilon_{\min} + T\ln\Big(1-e^{-n\lambda_T^2}\Big),
\end{equation}
where $\lambda_T^2=2\pi/(mT)$. Since $0<1-e^{-n\lambda_T^2}<1$, the logarithmic term is negative, ensuring $\mu_B(T)<\varepsilon_{\min}$, as required for the Bose distribution to remain finite. This construction preserves the same infrared cutoff used below $\Theta_c$, while replacing the collective phonon bath by a particle-like normal bath once the finite-size Bogoliubov criterion is no
longer satisfied.

\section{Phonon activation regime}
\label{sec:phonon_activation_window}

Before discussing the role of the impurity in Sec. IV, we now identify the finite-size momentum window in which thermally occupied phonons can dress the impurity. At zero temperature, the impurity is dressed by quantum Bogoliubov fluctuations of the condensate~\cite{StrongArdilla,Pastukhov2018}. At finite temperature, the finite infrared cutoff $k_{\min}$ sets a minimum Bogoliubov excitation energy, thereby controlling the thermal population of the lowest bath modes. This finite-size energy scale determines the onset of the following phonon-activation regimes:

\begin{itemize}
\item \textit{Activation temperature---}
The lowest accessible excitation has energy $E_{\min}(T)=E_{k_{\min}}(T)$. In the Bogoliubov regime, the low-momentum spectrum is phonon-like, $E_k(T)\simeq c(T)k$, with $c(T)=\sqrt{g_{BB}n_0(T)/m}$ being the sound velocity. Thermal occupation of the lowest mode becomes appreciable when the temperature is comparable to its excitation energy, which motivates the self-consistent definition of the activation temperature, $T_{\rm act}=E_{k_{\min}}(T_{\rm act})$. Since condensate depletion is negligible near this low-temperature onset for the parameters considered here, we approximate $T_{\rm act}\simeq E_{k_{\min}}(0) \simeq c(0)k_{\min}$. Thus, for $T < T_{act}$, thermal occupation of the lowest modes is exponentially suppressed, although zero-point Bogoliubov fluctuations remain present. The scale $T_{\rm act}$ therefore marks a smooth crossover to appreciable thermal population of the phonon modes rather than a sharp activation threshold.

\item \textit{Healing momentum scale---}
Once phonons become thermally active, one must still identify the
momentum range in which they are genuinely described by the infrared
Bogoliubov form. The lower bound is fixed by the finite-size cutoff,
$k>k_{\min}$. The upper bound is set by the healing scale
$k_\xi(T)=\xi^{-1}(T)$, with
$\xi(T)=\left(2mg_{BB}n_0(T)\right)^{-1/2}$ being the healing length. For momenta below this
scale, the Bogoliubov spectrum is approximately linear, whereas for
larger momenta the full particle-like curvature of the dispersion
becomes important.
\item \textit{Thermal momentum scale---}
The thermal momentum scale is set by the Bose occupation factor
$f_{\mathrm{B}}(E_k)$. In the phonon regime, the classical infrared
approximation
\begin{equation}
f_{\mathrm{B}}(E_k)
\simeq \frac{T}{E_k}
\simeq \frac{T}{c(T)k}
\label{eq:thermal_occupation}
\end{equation}
is parametrically accurate when $E_k(T) \ll T$. We therefore introduce
\begin{equation}
k_T(T)=\frac{T}{c(T)}
\end{equation}
as a crossover momentum. Modes with $k \ll k_T(T)$ are strongly
thermally occupied, whereas the infrared enhancement is progressively
lost for $k \gtrsim k_T(T)$. Similarly,
$k_{\xi}(T)=\xi^{-1}(T)$ should be understood as the crossover between
the approximately linear phonon regime, $k \ll k_{\xi}(T)$, and the
particle-like part of the Bogoliubov dispersion. The finite-size infrared contribution is therefore expected to be
concentrated in the momentum sector $k_{\min} < k < k_{*}(T),$
\begin{equation}
k_{*}(T)=\min\left[k_T(T),k_{\xi}(T)\right],
\label{eq:phonon_window}
\end{equation}
where $k_T(T)$ and $k_{\xi}(T)$ provide order-of-magnitude upper
crossover scales rather than sharp boundaries. This interval identifies
the momentum range in which thermally enhanced phonon contributions can
develop. The asymptotic approximations
$f_{\mathrm{B}}(E_k)\simeq T/E_k$ and $E_k\simeq c(T)k$ are most
accurate well inside this interval, away from its upper boundary.\\

\end{itemize}

\begin{figure}[!t]
    \centering
    \includegraphics[width = \linewidth]{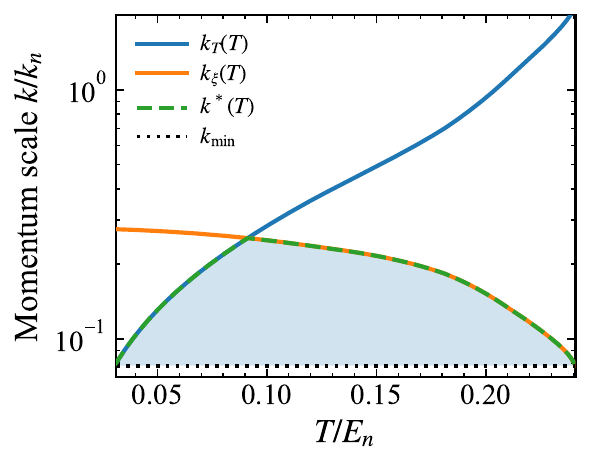}
    \caption{Temperature dependence of the characteristic momentum scales defining the infrared window: the thermal scale $k_T(T)$, the phonon (healing) scale $k_\xi(T)$, their minimum $k^*(T)=\min[k_T(T),k_\xi(T)]$, and the finite-size infrared cutoff $k_{\min}$. The shaded region indicates the momentum shell $k_{\min}<k<k^*(T)$ where the phonon infrared contribution develops. 
}
    \label{fig: IR_window}
\end{figure}

Figure~\ref{fig: IR_window} summarizes the characteristic momentum scales discussed for the parameters used in this work. At very low temperatures, $k_T(T)<k_{\min}$,  no thermally active phonon window is available and the lowest finite-size mode is not appreciably occupied. When $k_T(T)$ exceeds $k_{\min}$, a finite shell of thermally populated long-wavelength phonons emerges. At low and intermediate temperatures this shell is limited by $k_T(T)$, while at higher temperatures, still within the Bogoliubov regime, condensate depletion reduces $c(T)$ and shrinks the healing scale, so that $k_\xi(T)$ becomes the limiting cutoff. Thus the infrared phonon window first opens due to thermal activation and later narrows due to condensate depletion. This finite shell, rather than the formal $k\to0$ infrared limit of an infinite system, is the relevant low-energy sector controlling the thermal phonon dressing of the impurity.

\section{Bose polaron in 2D at finite temperature}
\label{sec:polaron}
We now turn our attention to the study of a single impurity immersed in this system. The full Hamiltonian is written as $\hat{H}=\hat{H}_B+\hat{H}_I$, with
\begin{equation}
    \hat{H}_I =
    \sum_{\mathbf{k}}\varepsilon_{\mathbf{k}}^{(I)}
    \hat{c}^{\dagger}_{\mathbf{k}}\hat{c}_{\mathbf{k}}
    +
    \frac{g}{\mathcal{A}}
    \sum_{\mathbf{k},\mathbf{k}',\mathbf{q}}
    \hat{a}^{\dagger}_{\mathbf{k}+\mathbf{q}}\,
    \hat{c}^{\dagger}_{\mathbf{k}'-\mathbf{q}}\,
    \hat{c}_{\mathbf{k}'}\,
    \hat{a}_{\mathbf{k}} .
\end{equation}
Here, $\hat{c}_{\mathbf{k}}$ and $\hat{c}^{\dagger}_{\mathbf{k}}$ are impurity annihilation and creation operators, respectively. The first term describes the kinetic energy of the impurity, with single-particle dispersion $\varepsilon_{\mathbf{k}}^{(I)}=k^2/2m_I$. The second term describes the impurity--boson interaction, which is modeled as a contact potential with coupling constant $g$. Since we consider the single-impurity limit, the quantum statistics of the impurity is irrelevant. We focus on the attractive polaron branch, corresponding to $g<0$. Throughout, we consider the equal-mass case $m_B=m_I=m$, so that $\varepsilon_{\mathbf{k}}^{(I)}=\varepsilon_{\mathbf{k}}$.

It is useful to contrast the attractive impurity considered here with the repulsive case studied in Ref.~\cite{Amelio2024}. For a density-coupled impurity, \(V_I({\bf r})=g\,n_B({\bf r})\), a repulsive interaction, \(g>0\), makes regions of depleted boson density, $n_B(\bf r)$, energetically favorable. Vortex cores therefore act as attractive potential wells for the impurity and can produce a distinct red-shifted impurity--vortex branch, as found in Ref.~\cite{Amelio2024}. In the attractive case studied here, \(g<0\), the situation is reversed, since the impurity lowers its energy by occupying regions of large boson density, especially in the strong-coupling regime. Vortices are therefore not expected to generate a separate low-energy impurity--vortex branch, but rather to contribute mainly to spectral broadening or weak energy-shift corrections. This is consistent with the attractive-interaction spectra of Ref.~\cite{Amelio2024}, where no additional impurity--vortex branch was resolved.

\subsection{Perturbation theory}

For weak impurity--boson coupling, the Hamiltonian of the system can be cast into a Fr\"ohlich--Bogoliubov form and treated perturbatively in the impurity--boson coupling strength $g$. At $T=0$, this approach has been shown to yield good agreement with quantum Monte Carlo calculations deep in the weak-coupling regime~\cite{StrongArdilla}. We start from the Fr\"ohlich Hamiltonian for a single impurity linearly coupled to the Bogoliubov modes of the Bose gas,

    \begin{align}
\hat{H}_{\rm Fr}
&=
\frac{\hat{\mathbf{p}}^{\,2}}{2m}
+\sum_{\mathbf{k}\neq 0}
E_{\mathbf{k}}\,
\hat{b}_{\mathbf{k}}^\dagger
\hat{b}_{\mathbf{k}}
+gn
\nonumber\\
&\quad+
\sum_{\mathbf{k}\neq 0}
V_{\mathbf{k}}
\left(
e^{i\mathbf{k}\cdot\hat{\mathbf{r}}}
\hat{b}_{\mathbf{k}}
+
e^{-i\mathbf{k}\cdot\hat{\mathbf{r}}}
\hat{b}_{\mathbf{k}}^\dagger
\right).
\label{eq:FrohlichHamiltonian}
\end{align}
The Fr\"ohlich vertex is $V_{\mathbf{k}}= g\sqrt{n_0}\left(u_{\mathbf{k}} - v_{\mathbf{k}}\right)$, where $u_{\mathbf{k}}$ and $v_{\mathbf{k}}$ are the Bogoliubov coherence factors, with $u_{\mathbf{k}}^2=\left[\left(\epsilon_{\mathbf{k}}+g_{BB}n_0\right) / E_{\mathbf{k}}+1\right] / 2$. For an impurity at zero momentum, the retarded Fröhlich self-energy is written as
\begin{equation}
\begin{aligned}
\Sigma_{\rm Fr}(\omega,T)
&=
gn + \int
\frac{d^2 k}{(2\pi)^2}
\, |V_{\mathbf{k}}(T)|^2
\\
&\quad \times
\left[
\frac{1+f_B(E_{\mathbf{k}})}
{\omega + i0^+ - E_{\mathbf{k}} - \varepsilon_{\mathbf{k}}}
+
\frac{f_B(E_{\mathbf{k}})}
{\omega + i0^+ + E_{\mathbf{k}} - \varepsilon_{\mathbf{k}}}
\right] .
\end{aligned}
\label{eq:FrohlichSelfEnergyFiniteT}
\end{equation}
Equation~\eqref{eq:FrohlichSelfEnergyFiniteT} contains the first-order Hartree shift, $\Sigma_{\rm Fr}^{(1)}$, and the second-order Fr\"ohlich contribution, $\Sigma_{\rm Fr}^{(2)}.$ The first-order term is proportional to the total boson density, $g[n_0(T)+n_{\rm ex}(T)]=gn$, where $n_{\rm ex}$ is the density of excited bosons (noncondensate), and is therefore independent of temperature at fixed density and coupling. The integral represents the condensate-assisted second-order processes in which the impurity scatters a boson into or out of the condensate through the emission or absorption of a single Bogoliubov excitation. Its temperature dependence is consequently both explicit, through the Bose occupation factors, and implicit, through the dependence of $V_{\mathbf{k}}(T)$ and $E_{\mathbf{k}}(T)$ on the condensate density. These are the diagrams conventionally identified as the Fr\"ohlich contribution in the weak-coupling perturbative treatment of the 3D Bose polaron ~\cite{Christensen2015,Jesper2017}.

The Fr\"ohlich sector does not contain direct scattering processes in which both the incoming and outgoing bosons belong to the noncondensed component. In microscopic perturbation theory, these appear as separate finite-temperature bubble diagrams and become increasingly important as the noncondensed population grows ~\cite{Jesper2017}. It also does not include the repeated impurity--boson scattering generated by a ladder resummation. Both extensions are incorporated later within the e-NSCT description, where the condensate-assisted and noncondensed contributions are constructed from the many-body $T-$matrix.

The occupation-induced thermal part of the second-order Fr\"ohlich self-energy is obtained from $\delta\Sigma_{\rm th}^{(2)}(\omega,T) = \Sigma_{\rm Fr}^{(2)}(\omega,T) - \Sigma_{\rm Fr}^{(2)}(\omega,0)$. In the lowest-temperature regime, we neglect the weak thermal variation of the condensate density, Fr\"ohlich vertex, and Bogoliubov parameters, so that this difference is determined by the terms explicitly proportional to the Bose distribution. The lower momentum bound is fixed by $k_{\min}$, while the upper bound $k_\xi$ restricts the calculation to the phonon-like part of the Bogoliubov spectrum. The corresponding energy scales are $E_{\min}=ck_{\min}$ and $E_\xi=ck_\xi$.

For frequencies $\Omega=\omega+i0^+$ sufficiently far from the absorption and emission thresholds, such that $E_{\mathbf{k}},\varepsilon_{\mathbf{k}}\ll|\Omega|$ throughout the relevant infrared window, the sum of the two denominators in the integral reduces to $2/\Omega$. The thermal correction then reduces to
\begin{equation}
\delta\Sigma^{(2)}_{\rm th}(\Omega,T)
\propto
{\frac{1}{\Omega}}
\int_{k_{\min}}^{k_\xi}
dk\,
k^2
f_B(ck).
\end{equation}
After the change of variables \(x=ck/T\), this gives
\begin{equation}
\delta\Sigma^{(2)}_{\rm th}(\Omega,T)
\simeq
\frac{g^2 n_0}{2\pi m c^4}
\frac{T^3}{\Omega}
\mathcal{I}_2
\left(
\frac{E_{\min}}{T},
\frac{E_\xi}{T}
\right),
\label{eq: T3 correction}
\end{equation}
where $\mathcal{I}_2(a,b)=
\int_a^b dx\,\frac{x^2}{e^x-1}$. The second-order Fr\"ohlich correction gives rise to a $T^3$ contribution to the polaron energy, $E_P$, once we identify it by the solution of $E_P = \mathrm{Re}\left[\Sigma_{\mathrm{Fr}}(E_P, T)\right]$, modulated by a finite-size phonon activation function. The $T^3$ scaling has the same phase-space scaling as the free energy of two-dimensional phonons, whereas the analogous contribution in 3D follows the expected $T^4$ behavior~\cite{Jesper2017}. The lower bound \(E_{\min}/T\) accounts for the activation of the lowest available phonon mode, while the upper bound \(E_\xi/T\) restricts the integral to the linear Bogoliubov regime. Equivalently, one could isolate the classical part of the thermally occupied phonon shell by using \(f_B(ck)\simeq T/(ck)\) for \(k<k_T\). Following the momentum-window analysis of Fig.~\ref{fig: IR_window}, this also gives a dominant activated contribution of \(T^3\) once \(k_T > k_{\min}\). In the expression above we keep the full Bose function in order to retain the finite-size correction and the smooth crossover associated with the opening and narrowing of the phonon window.

\begin{figure}[!t]
    \centering
    \includegraphics[width=\linewidth]{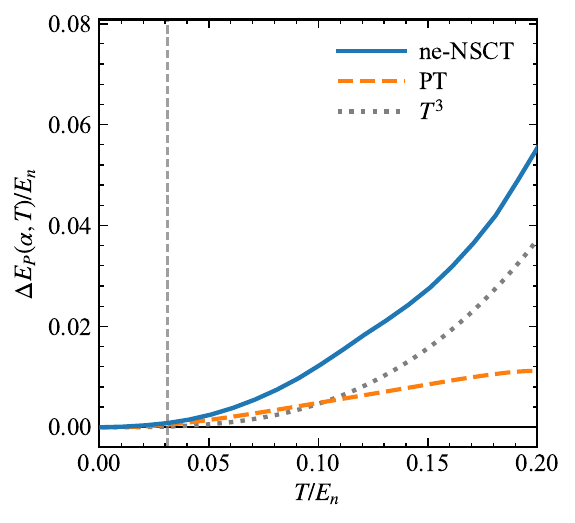}
    \caption{Weak-coupling energy shift $\Delta E_P(\alpha, T) $  for $\alpha = 10.0$. The perturbative Fr\"ohlich result is shown as perturbation theory (PT) and is compared with the ne-NSCT result. The gray dotted curve is a scaled $T^3$ guide representing the phonon-window phase-space dependence from Eq. \eqref{eq: T3 correction}. The vertical dashed line marks the finite-size phonon activation scale $T_{act}$.}
    \label{fig: PT}
\end{figure}

Figure~\ref{fig: PT} shows the finite-temperature variation of the attractive-polaron energy, as the difference between the polaron energy at finite and zero temperature, namely $\Delta E_P(\alpha,T)=E_P(\alpha,T)-E_P(\alpha,0)$, in the weak-coupling regime $\alpha=10.0$. The perturbative curve is obtained from the strict second-order Fr\"ohlich result, with the self-energy evaluated at $\omega=0$ in order to preserve the expansion consistently through second order in $g$ \cite{Jesper2017,Christensen2015}. Both perturbation theory and ne-NSCT (see next subsection) display a positive energy variation, meaning that the attractive branch moves upward and the impurity becomes progressively less dressed as temperature increases. The Fr\"ohlich diagrams already contain the emission and absorption of Bogoliubov excitations, including thermally populated modes, but these processes remain condensate-assisted. In the weak-coupling regime, they do not generate a net downward shift capable of overcoming the loss of condensate-assisted dressing. The ne-NSCT calculation shows the same trend, with a larger variation than strict perturbation theory because repeated impurity--boson scattering is resummed through the many-body $T$ matrix and a broader momentum structure is retained.

The vertical dashed line marks the activation temperature $T_{\rm act}$ set by the lowest accessible Bogoliubov mode. Below this scale, the thermal occupation of phonons is strongly suppressed and the energy varies only weakly. Above $T_{\rm act}$, low-energy collective modes become thermally populated and the  results develop the approximate $T^3$ dependence expected from the phase space of two-dimensional phonons, as in Eq. \ref{eq: T3 correction}. The fact that the perturbative and ne-NSCT curves remain positive even after phonon activation shows that, at weak coupling, the presence of thermally occupied Bogoliubov modes does not by itself imply dominant phonon-assisted dressing. Instead, the net response is still governed by the weakening of the condensate-assisted impurity dressing.

\subsection{Strong coupling: extended T-Matrix}

For the strong-coupling analysis, where the perturbative Fr\"ohlich treatment is no longer sufficient, repeated impurity--boson scattering processes must be resummed. We therefore use a non-self-consistent \(T\)-matrix approach, inspired by the e-NSCT formulation introduced for the 3D Bose polaron at finite temperature~\cite{Guenther2018}. The non-extended (ne-NSCT) approximation, shown in Fig.~\ref{fig: 1}(e), may be viewed as a nonperturbative extension of the condensate-assisted Fr\"ohlich processes discussed above, where instead of retaining only a single exchanged Bogoliubov excitation, it resums repeated impurity--boson scattering events in the ladder channel. At finite temperature, however, thermally excited bosons are also present in the bath, and the impurity can scatter from these non-condensed excitations or virtually scatter them back into the condensate. Such finite-temperature processes appear perturbatively as additional diagrams beyond the Fr\"ohlich contribution~\cite{Jesper2017}, and were incorporated in a ladder-resummed form in the e-NSCT scheme of Ref.~\cite{Guenther2018}. Here, we use this e-NSCT formulation for an impurity in a finite 2D Bose gas. Within this scheme, the impurity self-energy contains the usual condensate contribution, \(\Sigma_0(\mathbf{p},i\omega_n)\), and an additional noncondensate contribution, containing both quantum-depletion and thermal processes, \(\Sigma_1(\mathbf{p},i\omega_n)\), shown in Fig.~\ref{fig: 1}(e), which accounts for scattering from excitations outside the condensate. The corresponding extended scattering matrix, shown diagrammatically in Fig.~\ref{fig: 1}(d), reads

\begin{equation}
    \tilde{\Gamma}\left(\mathbf{p}, i \omega_n\right)=\frac{1}{g^{-1}-\Pi\left(\mathbf{p}, i \omega_n\right)-n_0(T) {G}_0\left(\mathbf{p}, i \omega_n\right)},
    \label{eq:scattering_extended}
\end{equation}
where $i\omega_n$ is a Matsubara frequency and  $G_0(\mathbf{p}, i\omega_n)$ is the bare impurity Green's function. Thus, the total self energy in the extended $T$-matrix formalism is $\Sigma = \Sigma_0 + \Sigma_1$, where $\Sigma_0(\mathbf{p}, i\omega_n) = n_0(T) \left[g^{-1} - \Pi(\mathbf{p}, i\omega_n)\right]^{-1}$, with $\Pi(\mathbf{p}, i\omega_n)$ being the pair propagator (bubble diagram), arising from the ne-NSCT contribution. The noncondensate contribution to the self-energy is

\begin{eqnarray}
\Sigma_1\left(\mathbf{p}, i \omega_n\right) & = & \int_{k_{min}}^{\infty} \frac{d^2 \mathbf{k}}{(2 \pi)^2}\Bigg[u^2_{\mathbf{k}} f_B(E_{\mathbf{k}}) \tilde{\Gamma}\left(\mathbf{k}+\mathbf{p}, i \omega_n+E_{\mathbf{k}}\right) \nonumber \\
& + &v^2_{\mathbf{k}}\left[1+f_B(E_{\mathbf{k}}) \right] \tilde{\Gamma}\left(\mathbf{k}+\mathbf{p}, i \omega_n-E_{\mathbf{k}}\right)\Bigg].
\label{eq:sigma1}
\end{eqnarray}
In order to proceed, one needs to calculate the pair propagator, which is written as 

\begin{eqnarray}
\Pi(\mathbf{p}, i\omega_n) & = & \int_{k_{min}}^{\infty} \frac{d^2 {\mathbf{q}}}{(2\pi)^2} \Bigg[ \frac{u^2_{\mathbf{q}}(1+f_B(E_{\mathbf{q}}))  }{i\omega_n - E_{\mathbf{q}} - \varepsilon_{\mathbf{p} - \mathbf{q}}}\nonumber \\ & + & \frac{v^2_{\mathbf{q}}  f_B(E_{\mathbf{q}})}{i\omega_n + E_{\mathbf{q}} - \varepsilon_{\mathbf{p} - \mathbf{q}}} \Bigg].
\label{eq:pair_prop}
\end{eqnarray}
This bubble diagram in Eq. \eqref{eq:pair_prop} is ultraviolet (UV) divergent in 2D, as it is at $T = 0$ \cite{Arturo2023}. We regularize this behavior by eliminating the bare impurity--boson coupling in favor of the vacuum two-body binding energy. In contrast with the 3D scattering problem, in 2D, an attractive short-range interaction always supports a two-body bound state. Denoting its binding-energy magnitude by $E_B > 0$, the bare coupling $g$ satisfies \cite{Randeria1990, Brunn2022, Pastukhov2018}

\begin{equation}
    \frac{1}{g}= -\int_0^{\Lambda} \frac{d^2 \mathbf{k}}{(2\pi)^2}\frac{1}{E_B+|\mathbf{k}|^2 / m},
    \label{eq: regularization}
\end{equation}
where $\Lambda$ is an auxiliary ultraviolet momentum cutoff. The lower limit here is $k = 0$ because this relation fixes the bare coupling through the complete thermodynamic-limit vacuum two-body problem. When Eq.~\eqref{eq: regularization} is combined with the self-energy contributions, the ultraviolet divergence of the momentum integrals is eliminated by the corresponding vacuum term. Consequently, all physical results can be expressed directly in terms of $E_B=|\epsilon_B| = 1/ma_{2D}^2$, which denotes the magnitude of the vacuum binding energy, while the physical energy of the two-body bound state is $-E_B$. This also introduces a direct dependence of the quasiparticle properties on the intrinsic 2D bound-state energy, which is exponentially dependent on $g$. From now on, we set the impurity-boson coupling through the dimensionless constant $\alpha = \ln{(k_n a_{2D})}$, which gives, in units of $E_n$, $E_B = 2e^{-2\alpha}$.\\

One important comment is that the vacuum binding energy defines the intrinsic two-body scale of the two-dimensional impurity–boson interaction, while the corresponding vacuum-dimer level is $-E_B$. This vacuum pole should be distinguished from the poles of the impurity Green’s function in the Bose medium. The many-body T-matrix entering the self-energy describes repeated impurity–boson scattering, but the resulting attractive-polaron pole additionally contains condensate-assisted, quantum-depletion, and thermal dressing processes. At large $\alpha$, weak interaction, the vacuum dimer is shallow and spatially extended, and the attractive branch remains predominantly impurity-like. As the attraction increases (small $\alpha$) and the dimer size becomes comparable to the interparticle spacing, the branch acquires increasing pair character and can, within the present two-body ladder approximation, be interpreted as a medium-dressed dimer. This interpretation does not imply the formation of a genuine impurity–multiboson bound cluster, since irreducible three-body and higher correlations are not included in e-NSCT.

We next isolate the infrared phonon contribution of the Bose gas and its effects on the impurity in this scheme. To expose this contribution, we separate the pair propagator into the occupation-independent contribution and its purely thermal correction, $\Pi(\mathbf{p},i\omega_n;T) = \Pi^{(0)}(\mathbf{p},i\omega_n) + \delta\Pi_T(\mathbf{p},i\omega_n),$ where $\Pi^{(0)}$ is obtained by setting $f_B(E_{\mathbf q})=0$ and $\delta\Pi_T$ contains the terms proportional to $f_B(E_{\mathbf{q}})$, therefore isolating the explicit thermal-occupation contribution. At zero external momentum, and inside the finite-size phonon window,  the Bogoliubov dispersion is linear, while the coherence
factors behave as $u_q^2+v_q^2 \simeq \frac{mc(T)}{q}$. For thermally occupied phonons and provided the energy denominators vary smoothly over the low-momentum shell, the leading thermal part of the pair propagator scales as
\begin{equation}
\delta\Pi_T^{\rm IR}
\sim
T
\int_{k_{\min}}^{k^*(T)}
\frac{dq}{q}.
\end{equation}
Thus,
\begin{equation}
\delta\Pi_T^{\rm IR}(\omega;T)
\sim
T\ln\left[\frac{k^*(T)}{k_{\min}}\right].
\label{eq:Pi_IR_log_main}
\end{equation}

This logarithm is the infrared many-body contribution generated by the 2D phase space of thermally occupied Bogoliubov phonons. In the
activated regime where $k_{\min}<k_T(T)<k_\xi(T),$ the upper limit is \(k^*(T)=k_T(T)=T/c(T)\), and Eq.~\eqref{eq:Pi_IR_log_main}
contains an explicit $T\ln T$ dependence, up to the slower temperature dependence of the sound velocity. At higher temperatures, but still below the breakdown of the condensate description, the condensate depletion reduces \(k_\xi(T)\). Once \(k_\xi(T)<k_T(T)\), the logarithm is instead cut off by the shrinking phonon window and becomes \(T\ln[k_\xi(T)/k_{\min}]\). Therefore the same infrared cutoff \(k_{\min}\) that activates the lowest phonons also controls the temperature range over which the logarithmic enhancement can develop. It is important to distinguish this infrared structure from the second-order Fr\"ohlich perturbative correction discussed above. In the Fr\"ohlich self-energy the thermal kernel contains the vertex factor \(|V_q|^2\propto q\), which softens the infrared behavior and leads instead to a phonon free-energy-like contribution proportional to \(T^3\), multiplied by a finite-size activation function. In the \(T\)-matrix formulation, by contrast, the pair propagator is repeatedly resummed in the impurity-boson scattering channel. The logarithmic thermal renormalization of \(\Pi\) therefore directly modifies the many-body scattering amplitude and can imprint a \(T\ln T\)-type dependence on the polaron energy in the regime where the upper infrared cutoff is \(k_T(T)\). This is a specifically 2D effect arising from the combination of Bogoliubov coherence factors, thermal phonon occupation, and the finite infrared momentum shell.

In the low-temperature, strong-coupling limit for a zero-momentum impurity with equal impurity and boson masses, the e-NSCT approach yields two quasiparticle branches that emerge symmetrically to leading order in the thermal population. In the corresponding three-dimensional calculation, the upper branch progressively loses spectral weight as the temperature increases~\cite{Guenther2018}. This splitting originates from resonant scattering with thermally occupied Bogoliubov modes and, in the present finite-size two-dimensional system, becomes appreciable only above the phonon-activation scale, since the infrared cutoff $k_{\min}$ suppresses the lowest-energy phonon spectral weight. Variational calculations show that the number of branches reflects the number of bath excitations retained in the ansatz, whereas the exact spectrum is expected to evolve into a broadened quasiparticle feature once higher-order correlations are included~\cite{Field2020}. Above $\Theta_c$, the condensate contribution vanishes and the extended scattering matrix reduces to the ordinary in-medium ladder vertex, with a single branch. The impurity is then dressed exclusively through repeated scattering from thermally occupied normal-state bosons.  Accordingly, the Bose gas is described in terms of thermal bath excitations rather than Bogoliubov phonons.

Finally,  after defining the self-energy and performing the usual analytical continuation $i\omega_n \rightarrow \omega + i\delta$, we can calculate the quasiparticle dispersion, $E_P(\alpha, T)$, as the solution of the self-consistent Dyson equation, written as $E_P(\alpha, T)-\operatorname{Re}\left[\Sigma\left(E_P, T\right)\right]=0$ \cite{Mahan2000}, where we focus on the zero-momentum impurity by setting $\mathbf{p} = 0$. Moreover, having calculated the self-energy, we also determine the spectral function, $A(\omega, T) = -2\mathrm{Im} G(\omega, T)$, where $G(\omega, T)$ is the dressed impurity Green's function. For the subsequent analysis, we perform a smooth interpolation on the $n_0(T)$ data from PIMC in order to have a continuous temperature dependence.

\section{Results}
\label{Sec: Temp_effects}

\begin{figure*}[!t]
    \centering
\includegraphics[width=\linewidth]{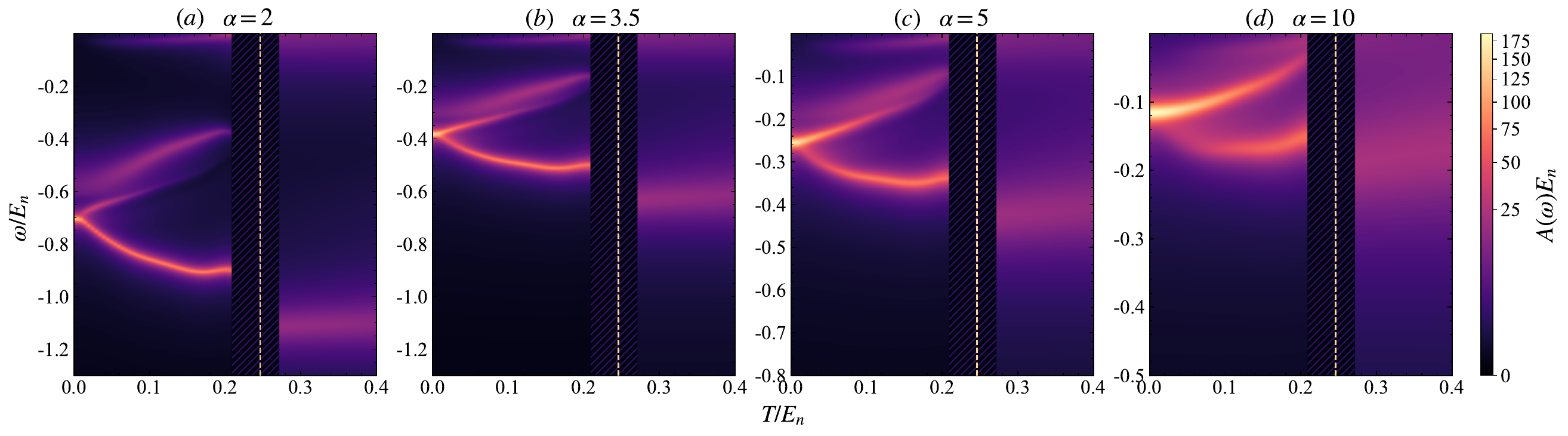}
    \caption{Zero-momentum impurity spectral function $A(\omega,T)$, showing the polaronic branch energies, $E_P$, as a function of temperature, calculated within the e-NSCT approximation, for progressively weaker couplings: (a) $\alpha=2.0$, (b) $\alpha=3.5$, (c) $\alpha=5.0$ and (d) $\alpha = 10.0$. The shaded interval marks the finite-size crossover around $\Theta_c$ (dashed white line), which is excluded from the quantitative analysis. We use a larger numerical width of $\delta/E_n = 0.05$ for $T > \Theta_c$ in the spectral line for clarity, while setting $\delta/E_n = 0.01$ for $T<\Theta_c$.}

    \label{fig: spectral_func}
\end{figure*}

\subsection{Temperature effects}

The zero-momentum spectral functions in Fig.~\ref{fig: spectral_func} resolve the polaronic branches and their redistribution of spectral weight within e-NSCT. At $T=0$, the spectrum contains a single attractive-polaron peak generated by condensate-assisted scattering and quantum fluctuations of the medium. Within the two-body ladder approximation, this state evolves from an impurity-like polaron at weak coupling toward a state with increasing impurity–boson pair character at stronger attraction \cite{Arturo2023}. At finite temperature, the resonant coupling between the condensate-assisted and noncondensed-boson contributions to the e-NSCT self-energy produces two polaronic branches. For the strongest coupling, $\alpha=2.0$, the lower branch remains continuously connected to the $T=0$ solution and most of the polaronic spectral peak intensity is carried by this branch throughout the condensed regime. As the interaction is weakened by increasing $\alpha$, the connection to the $T=0$ peak and the dominant spectral peak move to the upper branch, whereas the lower branch becomes resolved only after a sufficiently large thermal population develops and becomes progressively broader (see \hyperref[app:appendix B]{Appendix B} for the low-temperature analysis). The absence of arbitrarily soft modes below $k_{\min}$ introduces a finite activation scale in our system and delays the resolution of the splitting, which is less visible at weak coupling ~\cite{Guenther2018}. The two peaks found here should therefore be understood as the two quasiparticle poles resolved within the e-NSCT truncation, rather than as distinct many-body bound states.

The temperature evolution of the lower branch makes explicit the competition identified for the polaron energy. For $T<T_{\mathrm{act}}$, the modes above $k_{\min}$ remain weakly populated and the branch is nearly temperature independent, with the impurity dressing dominated by quantum fluctuations. Once $T_{\mathrm{act}}$ is crossed, thermally populated Bogoliubov modes enter the finite phonon window $k_{\min}<k<k^{*}(T)$, opening additional repeated-scattering channels and shifting the lower branch toward more negative energies. At higher temperatures, condensate depletion reduces $k_{\xi}(T)$ and closes this window from above. The additional thermal dressing then saturates and is opposed by the loss of condensate-assisted binding, leading to the weak upturn observed before the shaded crossover. These ingredients allow the impurity to respond directly to the thermally populated excitations of the bath and produce the nonmonotonic competition between condensate depletion and phonon-assisted dressing. In this sense, the strong-coupling polaron acts as a probe of the finite-temperature phonon fluctuations of the 2D bath. The remaining broad spectral intensity above the upper branch, including the structure near $\omega=0$, belongs to the continuum of impurity--boson scattering states and does not represent another polaron branch. The infrared cutoff does not create this continuum, but modifies its threshold and separation from the quasiparticle peaks. Above $\Theta_c$, only a broad attractive feature associated with scattering from thermal bosons remains and progressively approaches the free-impurity energy at high temperature.

The shaded region around $\Theta_c$ marks the finite-size crossover where the Bogoliubov description becomes marginal, rather than a sharp thermodynamic transition. As defined in Sec.~II, $\Theta_c$ is determined by $g_{\mathrm{BB}}n_0(\Theta_c)=\varepsilon_{\min}$, corresponding to the healing length becoming comparable to the infrared length scale set by the system size. In this regime, the momentum interval supporting collective phonon modes narrows, and the impurity energy becomes strongly sensitive to the reorganization of the low-energy bath. Above $\Theta_c$, we continue the calculation using a weakly interacting normal Bose gas with particle-like excitations. The results on the two sides of the shaded interval are obtained from different bath descriptions. Although the use of the smoothly interpolated condensate density $n_0(T)$ yields a continuous evolution of $E_P(T)$, neither approximation is quantitatively controlled throughout the crossover region. We therefore exclude this interval from the analysis, since the detailed behavior obtained there may depend on the specific matching prescription adopted between the two regimes. Developing a unified description capable of treating this crossover consistently remains an open problem and is required for a fully quantitative analysis of this region.

\subsection{Impurity-boson interaction effects}

\begin{figure}
    \centering
    \includegraphics[width=\linewidth]{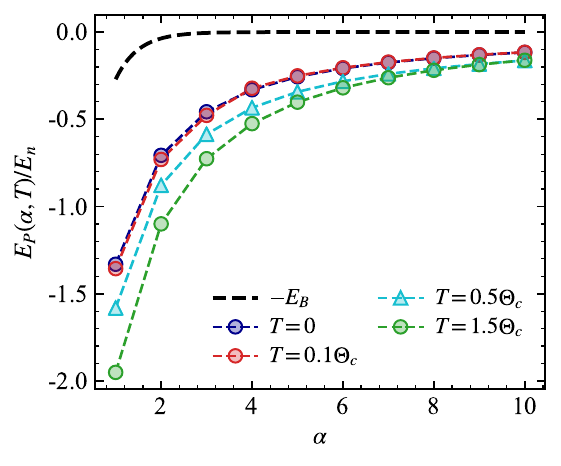}
    \caption{Polaron energies associated with the lowest branch (see Fig. \ref{fig: spectral_func}) within e-NSCT as a function of the impurity-boson interaction given by $\alpha$ for selected temperatures. For comparison, we also show the two-body bound state energy in vacuum, $-E_B$.}
    \label{fig:polaron_dimer_comparison}
\end{figure}

Figure~\ref{fig:polaron_dimer_comparison} compares the lowest attractive-polaron energy with the vacuum-dimer energy $-E_B$. At large $\alpha$, both energies approach zero. This visual convergence does not represent a molecular limit, since the bound state energy  vanishes exponentially with $\alpha$, whereas the weak-coupling polaron energy is governed by the much more slowly varying mean-field and many-body dressing scales. In this regime, the  lowest attractive-polaron branch is therefore better described as an impurity-like polaron generated through the two-body scattering channel.

As the attraction increases, the polaron energy moves substantially below the isolated dimer level. The difference represents the total medium renormalization within e-NSCT, arising from condensate-assisted scattering and quantum fluctuations. Temperature produces an additional shift, which contains the competition between condensate depletion and scattering from thermally occupied bath excitations. The increasingly negative displacement found at smaller $\alpha$, together with the persistence of a well-defined spectral peak, is consistent with the evolution toward a pair-dominated polaron that is strongly dressed by the Bose medium. Nevertheless, energy proximity or separation from $-E_B$ alone does not constitute a direct measurement of pair character, and the interpretation as a medium-dressed dimer remains specific to the two-body ladder structure retained by e-NSCT \cite{StrongArdilla}.

\section{Experimental considerations}
\label{Sec: experimental}

\subsection{Spectroscopic signatures of the finite-size crossover}
\label{subsec:probing_crossover}

A particularly appealing consequence of our results is that the attractive Bose polaron provides a sensitive probe of finite-temperature fluctuations in a confined 2D Bose gas. In the present finite-size setting, the impurity responds to the reorganization of the bath associated with condensate depletion and with the weakening of the  Bogoliubov phonon background. Our calculations suggest that the magnitude of the polaron energy variation across this crossover is controlled by the impurity--boson interaction parameter \(\alpha\), or equivalently by \(n a_{2D}^{2}\), which can be tuned experimentally through the density or the two-dimensional scattering length. Thus, the temperature-dependent polaron shift provides a sensitive
spectroscopic indicator of the finite-size Bogoliubov-breakdown crossover, rather than a direct or universal marker of a phase transition.

This signature could be probed using a dilute impurity fraction, with the impurity–boson interaction tuned through a Feshbach resonance. Measuring the temperature evolution of the polaron peak by radio-frequency spectroscopy~\cite{Shashi2014} or Ramsey interferometry~\cite{Etrych2025_1} should reveal a sizable shift of the impurity energy near the finite-size crossover, which is indeed observed at strong coupling for the 3D Bose polaron \cite{Yan2019}. For sufficiently large systems, the finite-size crossover studied here may occur in a temperature range close to that where BKT-related fluctuations become important. Since the present theory does not explicitly include vortices or superfluid-stiffness renormalization, the polaron shift should be interpreted as a probe of finite-size condensate depletion and thermal phonon dressing, rather than as a direct signature of BKT physics. Thus, while the attractive impurity primarily probes condensate depletion and thermally activated phonon dressing, its temperature-dependent energy shift may serve as an indirect spectroscopic marker of the BKT fluctuation regime. This provides an experimentally accessible window into finite-temperature fluctuations of 2D Bose gases~\cite{Schlederer2024}.

\subsection{Finite-size scales}

The finite system introduces two distinct temperature scales that organize the impurity response. The first and most relevant one is the phonon-activation temperature, $k_B T_{\mathrm{act}}\sim \hbar ck_{\min}=\frac{2\pi\hbar c}{L}$,
which marks the onset of appreciable thermal occupation of the lowest Bogoliubov mode allowed by the finite-size cutoff $k_{\min}=2\pi/L$. Below $T_{\mathrm{act}}$, this mode is exponentially weakly populated and the temperature dependence of the impurity dressing remains correspondingly small. Above $T_{\mathrm{act}}$, thermally occupied long-wavelength modes provide an additional dressing channel. Since $k_{\min}\propto L^{-1}$, the activation scale vanishes as $L\rightarrow\infty$. The second scale, $\Theta_c(L)$, is defined operationally by the condition $g_{B B} n_0\left(L, \Theta_c\right)=\varepsilon_{\min }(L)$ and identifies the upper boundary of the finite-size regime in which the lowest accessible bath modes remain interaction dominated and a condensate-based Bogoliubov description is meaningful. Functional renormalization-group and finite-size Monte Carlo studies show that the infrared cutoff allows a nonzero condensate fraction in finite two-dimensional systems, although this condensate vanishes with increasing system size at any fixed $T>0$ \cite{Floerchinger2009,Holzmann2007,HolzmannKrauth2008}. This should be distinguished from the superfluid response, which remains finite below the thermodynamic BKT transition~ \cite{Prokofev2001,Prokofev2002}. Accordingly, $\Theta_c(L)$ is a finite-size Bogoliubov-validity crossover rather than a thermodynamic condensation temperature. Since both $n_0(L, T)$ and $\varepsilon_{\min }(L)$ vanish as $L \rightarrow \infty$, its asymptotic behavior is determined by their relative finite-size scaling and cannot be inferred from the absence of true condensate order alone. For the long-distance estimator $n_0(L, T) \propto g_1(L / 2, T)$ employed here, standard BKT scaling suggests that $\Theta_c(L)$ tracks a BKT-related finite-size crossover; however, establishing its precise limiting behavior would require calculations for several system sizes. Thus, $T_{\mathrm{act}}$ controls the onset of phonon-assisted dressing, whereas $\Theta_c$ indicates the loss of a well-defined finite-size condensate and phonon background.

For the parameters considered here, the cutoff is fixed by $L=\sqrt{N/n}$, with $N=512$ bosons. For $^{87}\mathrm{Rb}$, using a boson-boson interaction $g_{BB}=0.5\hbar^2/m$, and the density range $n\simeq10\text{--}80~\mu{\rm m}^{-2}$ reported in Ref.~\cite{Ville2018}, one obtains characteristic system sizes $L\simeq2.5\text{--}7.2~\mu{\rm m}$ and activation temperatures in the approximate range $T_{\mathrm{act}}\simeq10\text{--}90~{\rm nK}$. The corresponding BKT scale is of order $T_{\mathrm{BKT}}\simeq45\text{--}420~{\rm nK}$, placing the activation crossover at approximately $20\text{--}25\%$ of $T_{\mathrm{BKT}}$. Although these systems are smaller than the homogeneous box employed in Ref.~\cite{Ville2018}, such length scales are compatible with the micron-scale shaping of quasi-two-dimensional density profiles using high-resolution optical potentials \cite{Zou2021}. The finite-size phonon-activation regime therefore lies within a realistic tens-of-nanokelvin temperature window.

Impurity radio-frequency spectroscopy provides access not only to the quasiparticle energy but also to the spectral weight and linewidth \cite{Jorgensen2016,Hu2016,Shashi2014}. Finite temperature modifies the polaron linewidth through scattering from thermally populated bath excitations, although the resulting response may broaden or narrow depending on the interaction strength and bath regime ~\cite{Etrych2025_2}. In the present work, however, we focus on the temperature dependence of the quasiparticle energy and do not provide a quantitative analysis of the spectral linewidth, since extracting it reliably would require a controlled calculation of the full spectral function throughout the crossover region. Within the Bogoliubov regime $T_{\mathrm{act}}<T<\Theta_c$, one nevertheless expects the linewidth to increase as thermally occupied phonons become available, together with a reduction of the quasiparticle coherence. A systematic calculation of this thermal broadening remains an important extension of the present analysis.

\begin{figure*}
    \centering
    \includegraphics[width=\linewidth]{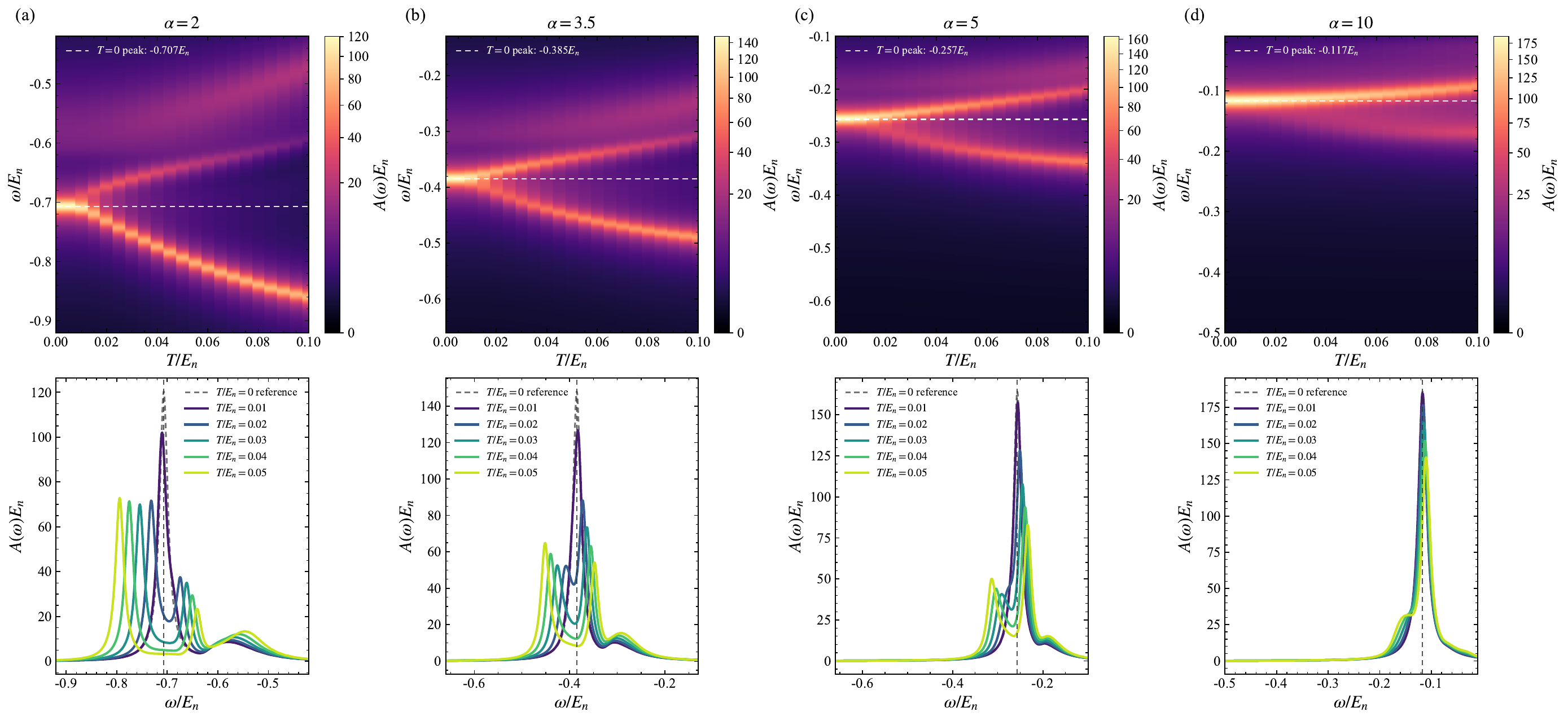}
    \caption{Spectral functions, $A(\omega, T)$, within the e-NSCT approximation, showing the polaronic branches at low temperature, $E_P(T)$, for progressively weaker couplings: (a) $\alpha = 2.0$, (b) $\alpha = 3.5$, (c) $\alpha = 5$, and (d) $\alpha = 10.0$. The upper panels show the heat maps of $A(\omega, T)$, with the horizontal white line representing the value of the energy peak at $T=0$. The lower panels show cuts of the spectral function heat map for the lowest temperatures, also highlighting the main peak at $ T = 0$ for reference with the black dashed line. }
    \label{fig: low-T spectral}
\end{figure*}
 
\section{Conclusions}
\label{Sec: conclusions}
In this work, we investigated the finite-temperature properties of an attractive impurity immersed in a finite two-dimensional Bose gas using PIMC, perturbation theory and an extended non-self-consistent $T$-matrix approach. This strategy provides a physically transparent description while being computationally less demanding than a fully numerical PIMC treatment across the entire parameter space. The finite system size introduced an infrared cutoff that supports a nonzero condensate fraction at low temperature and provided a controlled Bogoliubov background for impurity dressing. Within the Bogoliubov regime, we identified a phonon-activation scale $T_{\mathrm{act}}$, set by the lowest accessible Bogoliubov mode. For $T<T_{\mathrm{act}}$, thermal phonon dressing is strongly suppressed and the impurity energy varies only weakly with temperature. For $T_{\mathrm{act}}<T<\Theta_c$, thermally occupied long-wavelength phonons become available and compete with condensate depletion. The perturbative Fr\"ohlich theory includes thermal phonon emission and absorption, but it does not include direct scattering from noncondensed bosons or their repeated ladder resummation. Consequently, its thermal contribution is insufficient to produce the downward branch shift found within e-NSCT. In the nonperturbative regime, within e-NSCT, the interplay between condensate depletion and phonon dressing induces a nonmonotonic behavior of the polaron energy. Because the condensed and normal regimes are described using different bath approximations, the detailed behavior within the crossover interval around $\Theta_c$ is not a controlled quantitative prediction. Nevertheless, these findings indicate that the attractive Bose polaron can serve as a sensitive spectroscopic probe of finite-temperature fluctuations in confined two-dimensional Bose gases.\\

An interesting extension of this work would be to consider a finite density of fermionic impurities, enabling the study of boson-mediated impurity–impurity interactions, bipolaron formation~\cite{CamachoGuardian2018Bipolarons}, and pairing phenomena in two-dimensional quantum materials such as transition-metal dichalcogenide heterostructures~\cite{vonMilczewski2024Superconductivity}.

\section*{Acknowledgments}

Financial support from PNRR MUR project PE0000023-NQSTI is acknowledged. V.V. acknowledges financial support of PRIN 2022 (Prot. 20228YCYY7). This work has been supported by the Provincia Autonoma di Trento.

\newpage

\section*{Appendix}
\phantomsection
\subsection{Description of the PIMC method}
\label{app:appendix A}

In the present work, PIMC is used only to determine the temperature dependence of the finite-size long-distance coherence estimator entering the Bogoliubov description of the bosonic bath.
The PIMC method provides a systematically controllable approach for computing finite-temperature properties of interacting bosonic systems from the microscopic Hamiltonian. The method relies on a discretization of the imaginary-time interval $\beta=1/(k_B T)$ into a finite number of time slices, resulting in a Trotter decomposition of the thermal density matrix. This effectively maps the quantum many-body system onto a classical system of interacting polymer-like worldlines. Interactions are incorporated through the pair-product approximation \cite{Ceperley1995}, in which the short-time many-body density matrix is written as the free-particle density matrix multiplied by a product of two-body interaction correction factors. As in previous PIMC studies of two-dimensional systems~\cite{Carleo2013,Spada2024}, the interaction term is described by the spherical component of the two-body density matrix, which is justified in the weak-coupling regime, where it constitutes the leading-order contribution in the interaction strength~\cite{Gautier2021}. Sampling of the many-body configuration space is performed using the worm algorithm \cite{Boninsegni2006}, which enables efficient updates of bosonic exchange cycles and substantially improves the ergodicity of the simulation in the presence of quantum statistics. An additional advantage of the worm formalism is the direct sampling of off-diagonal configurations, allowing the evaluation of the normalized one-body correlation function $g_1(r)$~\cite{Boninsegni2005}. Additional details on the PIMC implementation can be found in Ref.~\cite{Spada2022}.

We performed simulations of $N=512$ identical bosons within a square box of side $L$ and periodic boundary conditions. For the present calculation, we retain the s-wave contribution to the two-body density matrix and use the dimensionless coupling $mg_{BB}/\hbar^2=0.5$. We selected temperatures spanning the expected thermodynamic-limit BKT transition region at the same density and boson–boson coupling. For each temperature, the number of imaginary-time slices is adjusted to maintain an adequate discretization of the thermal density matrix. Finer discretizations are employed at lower temperatures so that Trotter errors remain negligible compared with statistical uncertainties. Fig.~\ref{fig: 1}(b) shows the normalized one-body correlation function 
\begin{equation}
    g_1(r)=\frac{\left\langle \hat{\psi}^{\dagger}(\mathbf{r})\hat{\psi}(0)\right\rangle}{n},
\end{equation}
at the simulated temperatures. From these data, we define the finite-size long-distance coherence estimator $g_1(L/2)$ as also shown in Fig. \ref{fig: 1}(b). The correlation function is obtained from the histogram of relative head-tail separations sampled in the worm sector, providing a direct estimator of the long-distance one-body coherence of the finite system. Statistical uncertainties are computed from 12 independent simulations each with $2^{26}$ Monte Carlo steps after the initial thermalization phase.\\

\subsection{Low-temperature polaron branches}
\phantomsection
\label{app:appendix B}
We show the low-temperature analysis of the polaronic branch splitting in the spectral function $A(\omega ,T)$ within e-NSCT for selected impurity-boson couplings, $\alpha$, in Fig. \ref{fig: low-T spectral}. Here we can clearly see that only at the strongest coupling, $\alpha = 2.0$, is the lower branch directly connected with the $T = 0$ solution, carrying the dominant peak as the temperature is increased. As the impurity-boson interaction is weakened, the main $T = 0$ peak is now connected to the upper polaronic branch at finite temperature, while the splitting towards two branches becomes visible only for larger temperatures. We can also see the small bump above the upper branches related to the continuum of impurity-boson scattering states that, for a system with an IR cutoff given by $k_{min}$, becomes clearly visible since it does not overlap with the main quasiparticle peak, carrying a small spectral weight as a function of temperature. For the weakest coupling, $\alpha = 10.0$, the main peak overlaps with this continuum of states and the branch splitting becomes poorly visible, even for larger temperatures. Therefore, the lower polaronic branch that highlights the competition between condensate depletion and thermal phonon dressing is only clearly visible at strong coupling.

\phantomsection
\subsection{Continuum approximation and discrete-sum comparison}
\label{app:appendix C}

\begin{figure}
    \centering
    \includegraphics[width=\linewidth]{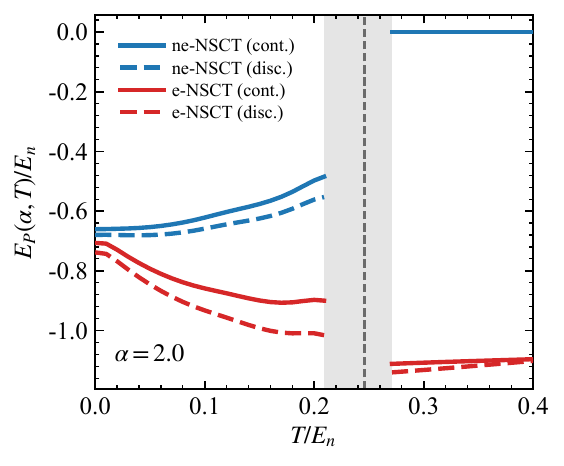}
    \caption{Comparison of the polaron energy at $\alpha=2.0$ obtained using the continuum formulation  (solid lines) and the corresponding shell-resolved discrete momentum sums (dashed lines), for ne-NSCT (blue) and e-NSCT (red). Both calculations use $N=512$ and the same bath parameters and condensate density.  The shaded region indicates the excluded crossover interval, while the vertical dashed line marks $\Theta_c$. For $T>\Theta_c$, the ne-NSCT energy vanishes because $n_0=0$, whereas the e-NSCT self-energy remains finite due to scattering from thermally occupied noncondensate bosons.}
\label{fig:continuum-discrete-comparison}
\end{figure}

In the calculations presented in the main text, finite-size effects are incorporated through the infrared scale $k_{min}=2\pi/L$, while momentum space is otherwise treated as continuous. Thus, our model does not describe a literal periodic box with a fully discrete spectrum. To assess the accuracy of this approximation, we compare the continuum integrals with a shell-resolved discrete representation constructed at the same system size,
\begin{equation}
\int_{k_{\min}}^\infty
\frac{d^2k}{(2\pi)^2}
\quad\longrightarrow\quad
\frac{1}{L^2}
\sum_{\substack{\mathbf{k}=(2\pi/L)(n_x,n_y)\ \mathbf{k}\neq0}}.
\end{equation}
The zero-momentum state is excluded from the physical pair propagator and self-energy sums, consistently with the continuum lower limit $k_{min}$. The vacuum subtraction defining the impurity--boson coupling retains its lower limit at $k=0$ in both representations. The ultraviolet cutoffs of the discrete sums are independently converged for the pair propagator and self-energies. At fixed $L$, increasing these cutoffs adds higher-momentum shells but does not reduce the spacing $\Delta k=2\pi/L$; consequently, the residual difference between the converged discrete and continuum results measures the effect of smoothing the finite-$L$ shell structure, rather than an ultraviolet truncation error.

Figure~\ref{fig:continuum-discrete-comparison} shows the resulting comparison for $\alpha=2.0$. Both implementations reproduce the same qualitative temperature dependence and the same distinction between the ne-NSCT and e-NSCT results. The discrete energies are more negative in the condensed regime because the first few momentum shells retain their exact degeneracies, while the continuum treatment replaces them by a smooth angularly averaged density of states. This effect is more pronounced in e-NSCT, for which the shell discretization enters both the pair propagator and the additional momentum sum in $\Sigma_1$. Nevertheless, both representations reproduce the phonon-activation scale, the nonmonotonic temperature evolution of the lowest e-NSCT branch, and the distinct behavior of each approximation. The comparison therefore establishes the qualitative robustness of the continuum IR-cutoff description.

\bibliography{sample}

\end{document}